# Support for public-key infrastructures in DNS

Marius Marian, Diana Berbecaru
Dipartimento di Automatica e Informatica
Politecnico di Torino
Corso Duca degli Abruzzi 24, 10129 Torino, Italy

**Abstract:** Traditionally, publicly available repositories of certificates offer the usual response to the problem of public key distribution. After issuing a public-key certificate a certification authority (CA) – in the frame of a particular *public-key infrastructure* (PKI) – will store and publish that certificate in a repository so that, at a later moment, end-users can search, find and retrieve public-key certificates. A known and still persisting drawback of this approach is that these repositories are not interconnected between each other on an Internet scale, therefore the search and retrieving of certificates on a wider scale turns out to be very difficult. In this scenario, end-users are supposed to know the Internet location of the repository before actually starting the procedure of search and retrieval. Currently, there are no means to perform automatic discovery of authoritative repositories for a particular certificate using as a search-key some information identifying an Internet entity. In this paper, we try to describe a different approach for solving the *key distribution* problem. This solution takes into account an already existing Internet-wide infrastructure: the *domain name system* (DNS).

## INTRODUCTION

In the field of Internet security, the major theoretical and technological breakthrough of the last century was the discovery of public-key cryptography (by M. Diffie and W. Hellman in 1976). With it, it became possible to have secure communication between Internet parties that had no a priori knowledge of each other. Public-key cryptography is based on the following principle: each party has a pair of keys (cryptographically related one with another) where one of the keys is made public, and the other is kept private. Typically, the public key is made available to everyone by means of a public-key certificate [1] (a.k.a. certificate, or digital certificate). This approach has the advantage of granting trust into the real owner of the public key and is accomplished by a trusted third party (TTP). The entity in charge of issuing certificates (i.e., the TTP) is called a certificate authority (CA). A CA will perform a set of verifications (including one on the identity of the owner of the public key) before digitally signing the certificate. Hence, a public-key certificate represents a signed (i.e. trustworthy) assertion regarding the owner's identity and the owner's public key.

## Why is Key Distribution an Issue in Public-Key Infrastructures?

The question titling this section may seem trivial at a quick look, however it hides a true problem when dealing with large-scale deployment of public-key technology within present-day Internet transactions. PKT-based security services require knowledge of public keys, and generally, a user of these services needs to obtain and validate certificates containing the required public keys. Currently, there are two types of globally available information retrieval systems that have utility: those designed for use by machines, and those designed for humans. The World-Wide Web is an example of an information retrieval system for humans, while DNS [2, 3] is an example of such a machine-oriented system.

Usually, one aspect that is often taken for granted, and therefore ignored in practice, is that before relying on public-key technology (PKT) for secure communication, Internet parties must in someway get possession of the public key (or the corresponding certificate) belonging to the other party. Let's assume for example the case of secure e-mail: in order to send confidential messages to his/her recipient(s), a user must obtain in some way the public key(s) of the recipients so that he/she can encrypt the message. Actually, the mechanism mentioned earlier relies on the user generating a symmetric key that will be used to encrypt the whole message, and afterwards the user will encrypt this symmetric key with the public key of each recipient. At receipt, each recipient will use his/her own private key to retrieve the encrypted symmetric key, and only then the recipient will be able, using this symmetric key, to decrypt the e-mail message. The case we wanted to emphasize is known as the problem of *public key distribution* (or certificate distribution).

Today, the majority of CAs allows users to retrieve public-key certificates and certificate revocation lists (CRL) by storing them in publicly available *repositories*. A comprehensive definition of a PKI repository is summarized in: "A repository is a system or a collection of distributed systems that stores certificates and certificate revocation lists (CRL), and serves as a means of distributing these certificates and CRLs to end entities".

From the point of view of certificate-using entities the problem of key/certificate distribution is divided into two major parts: *search* and *retrieval*. The first part of the problem is a bit more delicate since it involves the process of finding the location of a certificate when there are available either full or only partial information that specifies the subject of the certificate. For example, in the case of secure e-mail the user may possess only the e-mail addresses of the recipients, and therefore the user must identify and obtain the certificates of the recipients by means of just that information. The second part of the problem addresses the actual mechanism by which certificates can be fetched by end entities. Taking into account all these facts, we can see how critical are the certificate distribution mechanisms for PKI-based services and PKI operations. PKI repositories are clearly the solution, and a lot of efforts were made for standardizing and implementing solutions that could satisfy basic requirements such as the ability of users to search and retrieve certificates for a target identity. Unfortunately, the problem of repositories is that they are local, closely tied to the PKI community they belong and consequently the problem of "where to look for the certificate of a given entity" still persists. In the following paragraphs, we will describe the typical mechanisms currently available for end-entities when facing the problem of retrieving certificates and CRLs.

A *directory* is a specialized, distributed database that stores typed and ordered information. However, a directory is not a general-purpose relational database since it doesn't support complicated transactions or rollback schemes that are usual for traditional DBMS. For directories, only few operations are optimized such as reading, browsing and searching, and this is because the main goal of directories is to provide quick responses to high volumes of lookup and search operations. Additionally, the data typically stored in directories is changing infrequently while the directory updates are simple *all or nothing* changes. In order to support reliability and availability of service, directories have the ability to replicate information with short synchronization times between replicas. Taking into consideration all the advantages of a directory service, certification authorities have traditionally implemented their repositories using the directory model. The repository will be available to end entities by means of one or more directory servers.

Initially, X.509 certificates were thought to be stored in an X.500 directory. Due to technical, but mainly political difficulties encountered in trying to implement the full X.500 directory, there is currently no standard certificate distribution/storage mechanism. However, PKI repositories have been traditionally based on the *lightweight directory access protocol* (LDAP) [4, 5]. LDAP is much simpler than its precursor, the directory access protocol (DAP). LDAP is nothing more than an access protocol – compatible with the X.500 directory model – having one important advantage: it is *independent* of the particular technology employed by the underlying directory database. An LDAP directory stores information into a hierarchical manner, although the underlying database engine converts it back to flat format (that is, information is stored in rows of tables).

The management of an LDAP server requires a lot of effort for setting up schemas and storing certificates into the directory. Additionally, the ability of LDAP servers to store certificates depends entirely on the certificate content, since LDAP servers can only store certificates that comply with the server's scheme and expected attribute structure. Unless the certificates were specially designed to work with the server's scheme, the server would not be able to store them. Therefore, schemas need to be modified due to the inappropriate matching between the existing object classes and the ones required by some certificates. And this leads to other problems, such as the rebuilding of the entire directory information tree. Due to these facts, it is questionable if LDAP servers will ever be able to act as a general-purpose certificate store. In the same context, there should not be ignored the interoperability problems raised from having different schemes for different LDAP servers. Finally, it is worth mentioning that up to now the LDAP directories have failed to aggregate into a globally available infrastructure.

FTP (file transfer protocol) offers a viable solution for publishing and distribution of certificates and CRLs, being an attractive alternative to the traditional directory access protocols. A CA may choose to use an FTP server for publishing certificates and CRLs so that end entities can access this data by means of anonymous FTP. Below, there are depicted two uniform resource identifiers (URI) pinpointing user certificates and also one URI corresponding to a CRL, all of them based on an FTP distribution scheme:

*ftp://ftp.europki.org/italian_ca/certs/marius_marian.cer*
*ftp://ftp.europki.org/italian_ca/certs/ID235.cer*
*ftp://ftp.europki.org/italian_ca/crl/crl.crl*

FTP is one good choice for Internet users since it is widely deployed and also because anonymous FTP is accommodated by many firewalls. Therefore end entities in Internet can easily retrieve certificate and CRLs via FTP once they obtained the particular URI. The disadvantage of using an FTP-based repository service resides in the fact that certificates cannot be located when there is available only some partial information identifying the certificate-owner (e.g. the e-mail address).

HTTP represents another frequent solution for a PKI repository implementation. The idea is similar to the one previously described for FTP. A web server can represent for a CA an excellent means for publishing PKI-specific data. End entities can retrieve this data by means of HTTP-aware clients (e.g. Internet browsers). Examples of URI names pointing to certificates and CRLs are given below:

*http://www.europki.org/italian_ca/certs/marius_marian.cer*
*http://www.europki.org/italian_ca/certs/ID235.cer*
*http://www.europki.org/italian_ca/crl/crl.crl*

The HTTP-based repository service is practical when Internet users are already having the URI identifying the certificate. Its strong point consists in the fact that nowadays HTTP is largely deployed in Internet, and is well accommodated by firewalls. However, HTTP is not of great help in cases where users are trying to find a certificate having only some partial information regarding the owner of the certificate, for example an e-mail address.

The digital signature of a public-key certificate asserts its data content authenticity and integrity; hence, a public-key certificate doesn't need particular security protections. For this reason, a certificate can be distributed by means of a variety of other mechanisms including here transfers through non-trusted systems, or using non-secure protocols. For example, the S/MIME protocol enables an end entity to send its public-key certificate along with a S/MIME message.

Of course, the perspective of end entities waiting for S/MIME messages from all their peers in order to obtain their certificates just before being able to use public-key technology makes of little use such approaches.

Another mechanism for distributing certificates is to have certificates hard-coded into relying party's application (this usually happens in common used browsers such as MS Internet Explorer, Mozilla and Netscape Navigator). This approach has substantial drawbacks, adding and removing certificates is difficult and also requires the user intervention. In practice, frequently users don't have a complete understanding of all issues involved in certificate management. Moreover, the solution is incomplete since only a limited number of certificates can actually be hard-coded (the certificates of CAs) while the user – on its own – must somehow retrieve all other certificates. It becomes clear that the scalability of public-key technology is highly dependent on easily accessible PKI repositories.

## THE DOMAIN NAME SYSTEM AND ITS SUPPORT FOR KEY DISTRIBUTION

The Domain Name System (DNS) [2, 3] represents the set of protocols and services on a TCP/IP network that allows users of the network to use hierarchical user-friendly names when looking for other hosts instead of having to remember and use their IP addresses. This system is used almost by any other application and protocol that is involved in network communication (e.g., web browsing, ftp, telnet or other TCP/IP utilities on Internet). In the ISO/OSI hierarchy, DNS is placed at application level, even though its usage is transparent to the users that simply refer to names instead of IP addresses, and it can use either TCP or UDP as transport protocols. In practice, DNS can be seen as a distributed database of names. These names establish a logical tree structure called the *domain name space*. A name server may cache information about any part of the domain tree, but in general it has complete information about a specific part of the domain name space. This means the name server has authority for that subdomain of the name space – therefore it will be called *authoritative*. Resolvers are programs that extract the information from name servers in response to client requests. Usually, resolvers are mainly relying on UDP (since the DNS queries and responses are well-suited for this protocol), but TCP might be used whenever truncation of the returned data occurs.

A resource record (RR) represents the means by which the domain name system stores its data. Each RR may take one of the following two alternative forms:
  *domain_name [TTL] [CLASS] TYPE RDATA*
  *domain_name [CLASS] [TTL] TYPE RDATA*
The *domain_name* component of each RR specifies the owner of the data stored in the RDATA field. RRs are divided into classes, and each class denotes a type of network. Additionally, within each class of RRs, a type identifies the RR. Each type corresponds to a variety of data that DNS is able to store.

## Storing Public-Key Certificates in DNS

RFC 2538 [6] proposes a resource record useful for storing public-key certificates within the DNS denominated CERT. The format of the CERT RR (that is the RDATA field above) is depicted below:
  *Type KeyTag Algorithm Certificate/CRL*
The certificate type field is used for identifying the format of the certificate stored in the RR. Currently, DNS can store certificates conforming to X.509 standard, SPKI proposal, and PGP standard. Consequently, the *type* field can be represented either as an unsigned integer that corresponds to the particular certificate format, or as a corresponding mnemonic (e.g. PKIX, SPKI, PGP). The *key tag* field is identical with the one defined within the DNSSEC specification. In the DNSSEC perspective, the key tag is used to differentiate between multiple public keys when having to verify a DNSSEC signature. Each signature (SIG) RR has a key tag field that is unequivocally identifying the public key that was used for creating that signature. To reduce the computational effort involved in the verification process, the key tag field is used to efficiently select the appropriate public key. The same applies for the case of CERT RR. The algorithm for calculating the two-octet key tag is implemented by the following C-language function:

```
int compute_keytag (unsigned char *key, unsigned int keysize)
    {
       long int ac;
       unsigned int i;

       for (ac = 0, i = 0; i < keysize; ++i)
          ac += (i&1) ? key[i] : key[i]<<8;
       ac += (ac>>16) & 0xFFFF;

       return ac & 0xFFFF;
    }
```

The *algorithm* field is similar with the one pertaining to KEY and SIG resource records described in RFC 2535 [7]. A zero algorithm field represents the only exception; such a case indicates that the algorithm was not considered in the initial DNSSEC specification. The algorithm field is represented either as an unsigned integer or as a mnemonic in conformity with. The certificate is included in the RDATA component of the CERT RR in a base64-encoded form. The certificate content may be divided into white-space separated substrings. An example of a CERT RR containing an X.509v3 public-key certificate is given below (for space-saving reasons, it was given only a small part of the actual base64 encoding of the certificate):

marius.marian.polito.it 86400 IN CERT PKIX 30132 RSAMD5 (MIIFhzCCBG+gAwIBAgICA O4 wDQYJKoZIhvcNAQEFBQAwZTELMAkGA1UEBh MCSVQx HjAcBgNVBAoTFVBvbGl0ZWNuaWNvI GRpIFRvcmlubzE2MDQGA1UEAxMtUG9saXRlY25 pY28gZG…)

**Practical Experiments**

The purpose of these experiments was the implementation of a PKI repository starting from the following observations. Even though commonly used applications such as mail user agents (MUA) are PKI-enabled, in everyday practice e-mail messages are still exchanged without any means of protection. The main cause of this happening is the lack of an easy accessible, globally available repository. A secondary cause is the unawareness of people using the technology. While the second cause is difficult to eliminate, we believe that for the technology-related cause an easy solution can be provided. Additionally, IPsec is today the most common way in which security can be achieved at network layer. In practice, IPsec has been widely deployed to implement virtual private networks (VPN). In order to have a secure exchange of packets at IP layer, both the sending and the receiving devices must share a cryptographic key. The key management protocols involved in the IPsec's key-exchange scheme can take advantage of a DNS-based repository.

The practical experiments dedicated to testing the DNS support for PKI services were performed in the frame of EuroPKI [8] public-key infrastructure. The EuroPKI Certification Authority is a non-profit organization established to create and develop a pan-European public-key infrastructure. Politecnico di Torino hosts the Root CA of the EuroPKI project. The experiments aimed to implement a DNS-based PKI repository for the certification authority of Politecnico di Torino (POLITO). For our experiments we have used Bind v9.1.2 [9] a popular, open-source DNS implementation. The master nameserver was installed on a Sun Ultra 5 workstation (Solaris 2.7) with a UltraSPARC-II CPU at 333 MHz and 256 MB RAM memory, while the secondary nameserver was installed on a Intel PIII 1 GHz Linux box with 256 MB of RAM. The zone files of these nameservers included all valid public-key certificates issued by POLITO CA. The experiment involved the most common types of certificates issued by POLITO CA:
o  *personal* certificates are used to bind a public key to the identity of an individual, and
o  *server* certificates are used to bind a public key to the identity of a network node or of a network service.

These X.509v3 certificates issued by POLITO CA are not minimal certificates, since they include a significant set of certificate extensions. The average size of the DER-encoded POLITO CA certificates is around 1340 bytes for server certificates, and approximately 1400 bytes for personal certificates. It is worth mentioning that all these certificates contain 1024-bit public keys.

Previously, the certificates of POLITO CA were distributed to relying parties by means of an LDAP-based repository and also by means of a HTTP-based repository. The personal certificates issued by POLITO CA are mainly used for securing services such as e-mail, and also for client authentication within PKI-aware applications, such as SSL-telnet and SSL-ftp.

**Proposed Naming Scheme**

The CERT resource record allows mapping of public-key certificates to domain names. Obviously, this mapping creates the opportunity of transforming the DNS into a globally available PKI repository. The primary requirement of DNS resides in its need for having some domain name associated with each entry. The standard requires that CERT RRs should be stored under a domain name related to their subject, that is the identity of the entity intended to control the private key corresponding to the certified public key. Translating the subject's distinguished name into a domain name is frequently a delicate and rather difficult problem. It is impossible to have a unique and definitive solution to this problem. Therefore, the standard provides a set of alternative solutions that could meet the above requirement. These solutions should be used in practice respecting the following order:
o  If a domain name is used for the identification of the certificate's subject, then that domain name should be used.
o  If a domain name is not included but an IP address is included, then the translation of that IP address into the appropriate inverse domain name should be used.
o  If neither of the above is used but a URI containing a domain name is present, then that domain name should be used.
o  If none of the above is present but a character string name specifying the subject's e-mail address is included, then the standard translation of the subject's e-mail address into a domain name should be used.
o  If none of the above applies, then the distinguished name (DN) should be mapped into a domain name as specified in RFC 2247 [10].

Taking for example the personal certificates issued by POLITO CA, it can be observed that each certificate contains a *SubjectAltName* extension. This extension specifies that an alternative name for the certified subject is his/her e-mail address. Starting with 2002, the internal regulation of Politecnico di Torino imposes for each individual affiliated to this institution mailing addresses in the form of: *first_name.last_name@polito.it*. The RFC 822 [11] format used for mailing addresses can be easily

translated into a domain name just like in the following example:

 marius.marian@polito.it ? marius.marian.polito.it

Similarly, POLITO server certificates have the *SubjectAltName* extension indicating the *dNSName* of the host on which the server runs. The naming scheme used for storing certificates in our DNS-based repository will obey the following two rules:
- In case of a personal certificate, it will be used the domain name corresponding to the standard translation of the individual's e-mail address (stored in the *SubjectAltName* extension).
- In case of a server certificate, it will be used the domain name stored in the *SubjectAltName* extension.

**Administration of Certificates**

The default policy is that all certificate of POLITO CA can be published in the DNS-based repository. However, certificate subscribers are questioned (at application time) if there are any privacy requirements impeding storage of their certificates in DNS. For publishing certificates into the DNS-based repository we have developed a set of software tools. These tools are used by the CA operator for the administration of the repository. The operations that must be performed for each newly issued certificate are:
1. Search for the *Subject* field within the certificate content. Extract the value stored under this field corresponding to either an e-mail address or a domain name. In case of multiple choices, it will be applied – respecting the specific priorities – the CERT naming rules mentioned earlier.
2. If the above process failed, search for the *SubjectAltName* extension within the certificate content. Extract the value stored under this extension corresponding to either an e-mail address or a domain name. In case of multiple choices, it will be applied – respecting the specific priorities – the rules mentioned earlier.
3. Once a domain name is determined, it will be created a CERT resource record for each newly issued certificate. For each CERT resource record, it will be determined the appropriate values of the necessary fields: TTL, class, and type. Additionally, it will be computed the RDATA fields: certificate type (usually PKIX certificates), the key tag and the algorithm type. Finally, the base64-encoding will be used for storing the actual certificate.
4. Once a new RR is created the CA operator can upload this RR into the DNS database. After the DNS database was updated with the latest information, the nameserver will be signaled to reload its data.

**Scalability and Performance**

The transition to a new model for PKI-repository requires a careful analysis and one of the things to be considered before moving towards an innovative model is its scalability. Since most PKI repositories are built today on LDAP technology we found reasonable to compare the LDAP-based repositories with the DNS-based approach. The tests were performed using open-source implementations of LDAP and DNS. For LDAP the immediate choice was the OpenLDAP [12] implementation, whilst for DNS we have chosen the well-known implementation of BIND provided by the Internet Software Consortium.

One PKI problem is that mapping from a X.500 name to a different name space often becomes extremely difficult. Since Internet communication today is naturally expressed in terms of DNS names, it is reasonable to have DNS-based repositories for public-key certificates if we want to provide security features based on PKI technology to Internet transactions.

The first disadvantage of LDAP compared with DNS is that, currently, it fails the test of deployment on a global scale. Moreover, locating a certificate can turn to be extremely difficult if a relying party does not know which is the authoritative LDAP server that can answer a potential search request. Usually, a relying party searching for another entity's certificate is holding only a partial set of information identifying the target entity: an e-mail address or simply a host name. Discovering authorities in DNS was a design goal and Internet works today also thanks to this feature. Resolvers are able to parse the entire DNS tree (using referrals from intermediate nameservers) in order to find an authoritative name server they can interrogate.

Nowadays, LDAP server implementations allow administrators to configure referrals to other LDAP servers in cases where requests arrive for data outside the authoritative domain. However, given that so far it wasn't actually implemented a global LDAP infrastructure (as is the case with DNS) – and, more importantly, there are no perspectives in the near future for a global standard meant to link LDAP servers to each other – the usage of referrals in LDAP is of little use. In other words, an end entity - having only a small piece of information about its peer – would have problems in determining the authoritative LDAP server that could provide the peer's certificate. Consequently, discovery of authorities is difficult in LDAP, and it can't be done in a dynamic and simple way (as it is currently done in DNS). Moreover, partitioning the directory information tree is possible in LDAP, even though it is a lot harder than in DNS, and this operation always requires the presence of a master server. But, partitioning the tree usually results in a non-uniform distribution of data, and every query still has to go through the root of the tree. In this way, searching the tree will always be limited by the performance of any one single directory server. All these facts put in doubt the current scalability potential of LDAP.

An LDAP query will always require a TCP connection, thus the TCP protocol overhead will always be present with its inherent latency. On the other hand, DNS is capable to operate on both TCP and UDP. The advantage of using DNS over UDP consists in the fact that a client will always make one query and will receive one response (be it a referral or an authoritative answer). It was observed that LDAP requires one round trip to set up the connection followed by two round trips for sending a client bind request and receiving the server bind response. Then, the actual lookup query requires another round trip. When answering a search query, the LDAP protocol assembles the response data in so-called search entry packets. There can be zero or more such packets depending on how many LDAP entries have matched the lookup filter. However the status of this lookup transaction will be always sent in a search response packet. Immediately after the search response was delivered, the client usually closes the TCP connection by means of an LDAP unbind request. The fact is that one TCP connection needs a minimum of 5 packets (usually 6) for setup and tear down, excluding data packets, thus requiring at least 3 round trips on top of the one for the original UDP query! During the practical tests, it was observed that a DNS-based query is taking fewer round trips than an LDAP-based query even when the DNS-based query was to be retried via TCP. However a DNS resolver can be easily instructed to start a connection directly via TCP avoiding thus the sometimes-useless round trip involved in the initial UDP transaction.

Traditionally, the size of DNS messages on UDP was limited to a maximum size of 512 bytes. The 512 bytes limit was imposed in the first place to reduce the probability of fragmentation of DNS responses. Lately, efforts have been made to extend the performance of DNS nameservers in order to support DNS messages greater than 512 bytes. The DNS extension mechanism EDNS0 [13] allows resolvers to inform nameservers that they are able to process DNS responses larger than 512 bytes. Thus, if the expected answer is between 512 octets and the maximum size that the client can accept (that is the maximum transfer unit of the client's stack), the additional overhead of a TCP connection can be avoided. Using this extension mechanism in a series of tests, we have noticed that DNS is able to use without problems UDP for DNS messages up to 4096 bytes on an Ethernet network (where the maximum transmission unit is 1500 bytes). Of course, different OS stacks impose different upper limits for the UDP datagrams those stacks can handle and reassemble.

Both TCP and UDP have their specific overhead (20 bytes for TCP and 8 bytes for UDP). Taking into account these values, the maximum payload for an Ethernet packet results to be of 1460 bytes when using TCP/IP, and 1472 bytes via UDP/IP. If we correlate the maximum payload with the sizes of the certificates described earlier (POLITO CA certificates range between 1340 – 1400 bytes), we can easily see the advantage of using DNS for retrieval of certificates. Moreover, since the majority of OS stacks today are able to reassemble UDP packets up to 4 KB then there is not a problem for applications to use DNS for locating and fetching public-key certificates whenever the DNS extension mechanism (EDNS0) is present.

Another interesting aspect regards the support for the two protocols in the operating systems available today. On one hand, for DNS we have universal OS support, whilst for LDAP few OS are supporting it now, consequently patches or/and updates are necessary on client platforms.

The DNS-based approach presented in this paper has two important issues that require careful attention for implementation: first, the transition from X.500 distinguished names to DNS names, is not always as straightforward as that encountered in our experiments. Second, firewalls are known to interfere with the UDP protocol (even though it is assumed that the DNS traffic will be allowed) consequently, it may happen that, sometimes, the DNS queries and responses will be blocked.

## CONCLUSIONS

One serious obstacle to the availability of public-key cryptography everywhere and every time is the lack of a worldwide, easy-accessible repository for digital certificates. DNS provides a unique opportunity for PKIs: to take advantage of a system that is already deployed on a global scale, and which conforms exactly with the way in which people in Internet communicate the name of the host to whom they want to connect, or the name of the person to whom they want to send a message. The latest security extensions of the DNS protocol support a new, encouraging perspective for public-key technology: extended DNS may act as a global PKI repository. This potential can be used by a variety of PKI-aware protocols (such as S/MIME, and IPsec).